\documentstyle[14lomcon,cite,epsfig]{article}

\begin{document}

\title{ELECTROWEAK PHYSICS AND SEARCHES FOR NEW PHYSICS AT HERA}

\author{ U.Schneekloth \footnote{e-mail: uwe.schneekloth@desy.de} }

\address{Deutsches Elektronen-Synchrotron DESY \\ 
Hamburg, Germany}


\maketitle\abstracts{ Recent results from the H1 and ZEUS experiments
are reported on electroweak physics and on searches for new physics. 
All results are in good agreement with the Standard Model.}

\section{Introduction}

High energy electron- (positron-) proton collisions at the HERA collider,
colliding 27.5\,GeV electrons (or positrons) on  920\,GeV  protons,
provide a unique opportunity for studying electroweak physics and for
searches for physics beyond the Standard  Model. 
Extensive studies have been performed by the H1 and ZEUS collaborations
with the final data sets, corresponding to an integrated luminosity of 
about 0.5 fb$^{-1}$ per experiment.
Recently, significant improvements have been achieved by combining the 
results of both experiments. 
A summary of electroweak studies and searches for new physics 
is presented in this paper.

\section{Evidence for Electroweak Unification}

One of the main goals of the HERA physics program has been the precise
measurement of the differential cross sections in neutral  (NC) 
and charged current (CC) deep inelastic scattering (DIS). 
These measurements have not only been of great importance for the 
understanding of the quark and gluon content of the proton, 
but have also provided a basis for electroweak (EW) studies~\cite{Cooper}.
%
%
\begin{figure} [h]
\vspace{-5.0mm}
\centerline{\psfig{figure=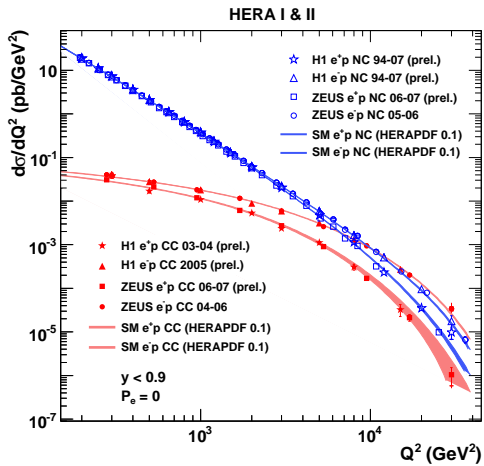,width=6.7cm}}
\vspace{-2.0mm}
\caption{\label{f:NCCC}  {\it  Differential cross section for NC and CC 
	$e p$  scattering, as measured by H1 and ZEUS, compared to the SM 
	expectations using the HERAPDF parametrization of the proton
        parton distribution functions.  \hspace{\fill}
 }}
\vspace{-2.0mm}
\end{figure}

Figure~\ref{f:NCCC} shows the single differential NC and CC $e^+p$ and $e^-p$
cross sections  measured by H1~\cite{H1_NC,H1_CC} and ZEUS~\cite{ZEUS_NCCC}
for $Q^2 >$ 200\,GeV$^2$ as a function of the momentum transfer $Q^2$.
The NC data show a $1/Q^4$ behavior due to the electromagnetic current,
whereas the CC cross section is proportional to  
$1/(Q^{2}+M_{W}^{2})^2$, which gives a much less steeper cross section 
decrease as a function of $Q^2$, since the propagator  term includes the 
$W$ mass. 
At high $Q^2$, $Q^2 > M^2_W$, both NC and CC are mediated by a unified 
electroweak current and both cross sections are of comparable size. 
There is excellent agreement with the SM predictions 
over 7 (4) orders of magnitude for NC (CC) scattering.

\section{Neutral Current Cross Sections }

\begin{center}
\begin{figure}[h]
  \unitlength 1cm
  \begin{minipage}{12cm}
    \begin{minipage}{5.9cm}
     \vspace{-10mm}
      \begin{picture}(5.9,5.9)
        \leavevmode
        \epsfig{file=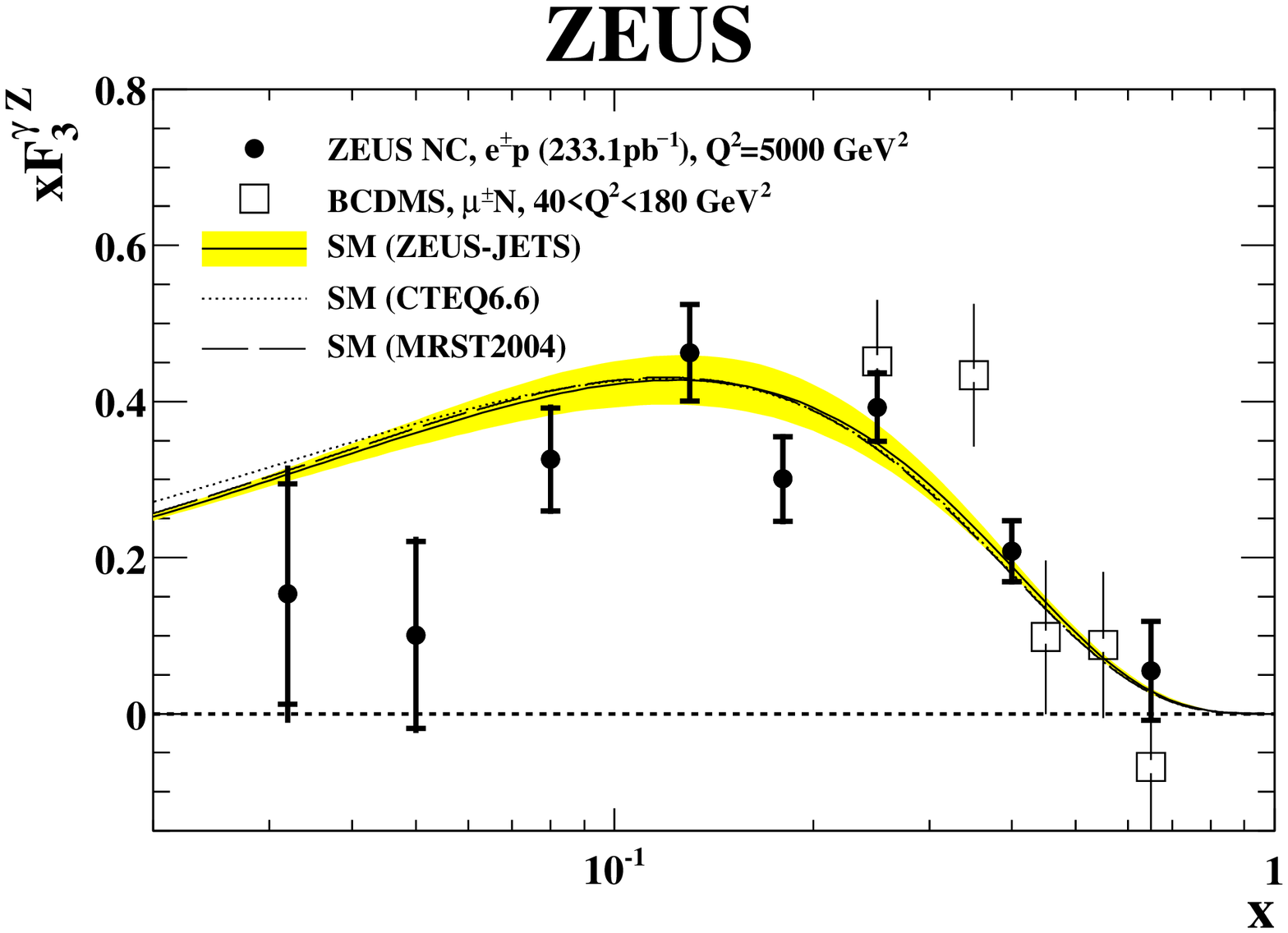,width=1.00\textwidth}
      \end{picture}
      \vspace{0.2mm}
     \caption{ {\it The structure function $xF_3^{\gamma Z}$ 
             extrapolated to a single $Q^2$ value of 5\,000\,GeV$^2$, plotted 
             as a function of $x$.  \hspace{\fill}
	     }} 
     \label{f:xF3}
    \end{minipage}
    \hfill
    \begin{minipage}{5.9cm}
    \vspace{7.0mm}
      \begin{picture}(5.9,5.9)
          \leavevmode
        \epsfig{file=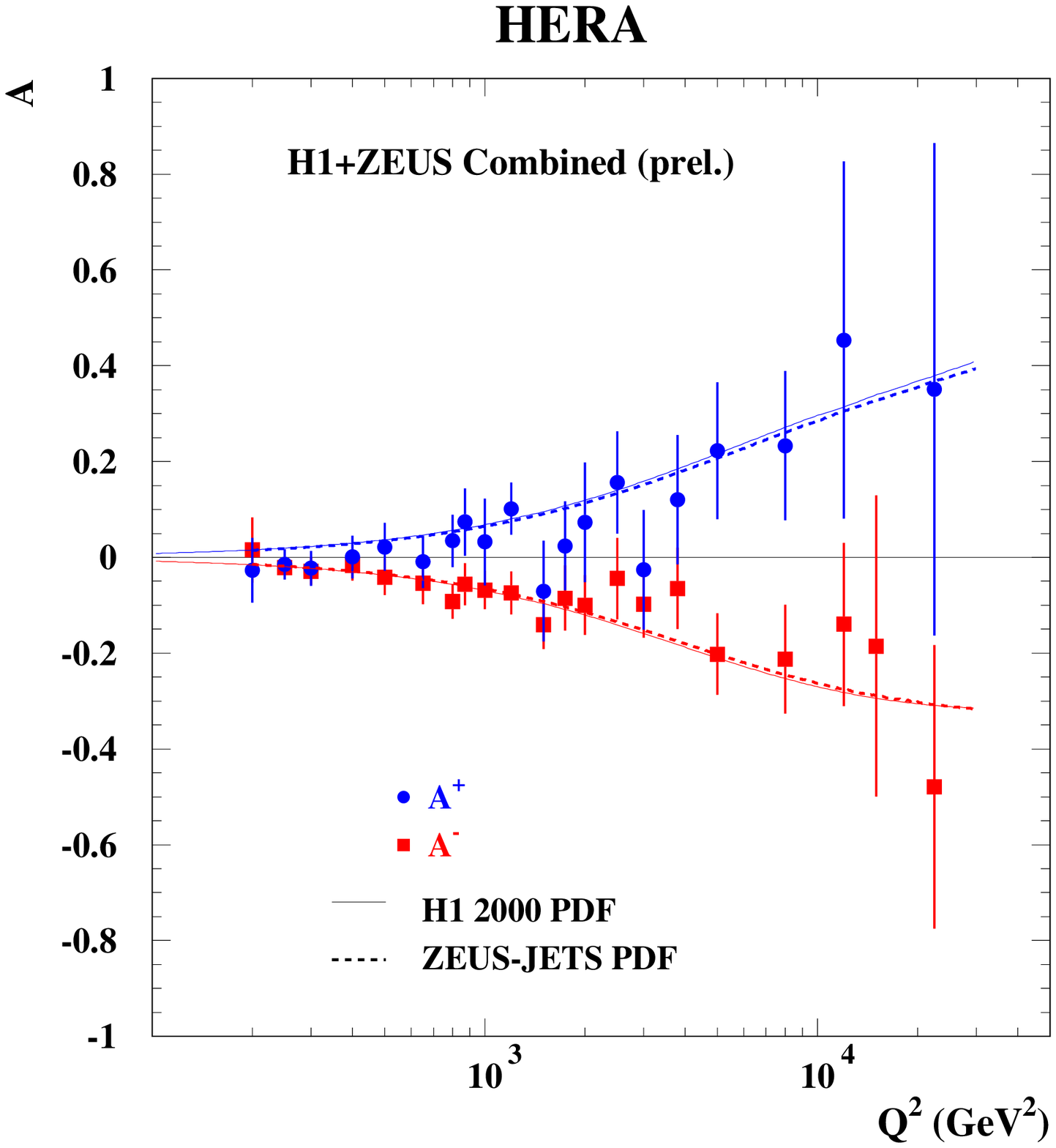,width=1.00\textwidth}
      \end{picture}
      \begin{center}
      \end{center}
      \vspace{-20.0mm}
      \caption{ {\it The polarization asymmetry $A$ plotted as a 
              function $Q^2$.  \hspace{\fill}
	   }} 
     \label{f:PolA}
    \end{minipage}
  \end{minipage}
\end{figure}
\end{center}

The unpolarized NC cross section has been measured over a large 
range of $Q^2$ (200 - 30\,000\,GeV$^2$) for $e^- p$ and $e^+ p$ DIS. 
At high $Q^2$, the $e^- p$ cross section is significantly larger 
than the $e^+ p$ cross section. 
This charge asymmetry can be exploited to measure the interference
structure function $x F_3^{\gamma Z}$:
\begin{displaymath}
x F_3^{\gamma Z}  
 \simeq x \tilde F_3 \frac{(Q^2+M_Z^2)}{\alpha_e \kappa Q^2}, \hspace{5mm}
 x \tilde F_3 =  
       \frac{Y_+}{2Y_-} (\tilde \sigma^{e^-p} - \tilde \sigma^{e^+p}).
\end{displaymath}
Since $x F_3^{\gamma Z}$ has little dependence on $Q^2$, 
the measurements from 
$1\,500 < Q^2 < 30\,000$\,GeV$^2$ were  extrapolated to 5\,000\,GeV$^2$ and
then averaged to obtain higher statistical significance. 
Figure~\ref{f:xF3} shows $x F_3^{\gamma Z}$ measured at 
$Q^2 = 5\,000$\,GeV$^2$~\cite{ZEUS_xF3}.
It is  well described by the SM prediction.

%

A direct measure of EW effects are the charge dependent polarization
asymmetries of the NC cross sections, which are now accessible using the
HERA~II data. 
The cross section asymmetries $A^\pm$, as defined below, can be used
to measure to a good approximation the structure function ratio, 
which is proportional to the product $a_e v_q$,
where $a_e (v_q)$ is the axial (vector) coupling of the electron 
(quark $q$) to the $Z$ boson, 
and thus gives a direct measure of parity violation.
\begin{displaymath}
A^\pm =  \frac{2}{P_R-P_L}  \cdot
    \frac{\sigma^\pm(P_R)-\sigma^\pm(P_L)}{\sigma^\pm(P_R)+\sigma^\pm(P_L)}
   \simeq \mp \kappa a_e \frac{F_2^{\gamma Z}}{F_2} \propto a_e v_q, 
\end{displaymath}
where $P_R$ ($P_L$) is the right (left) handed lepton 
beam polarization. 

The asymmetries obtained from the combined H1 and ZEUS data are
shown in Fig.~\ref{f:PolA}~\cite{H1_ZEUS_PolA} and are well described 
by the SM predictions
as obtained from the H1 and ZEUS QCD fits.
The data demonstrate parity violation at very small distances, 
down to $10^{-18}$\,m.

\section{Polarized Charged Current Cross Sections}

\begin{figure} [h]
\centerline{\epsfig{figure=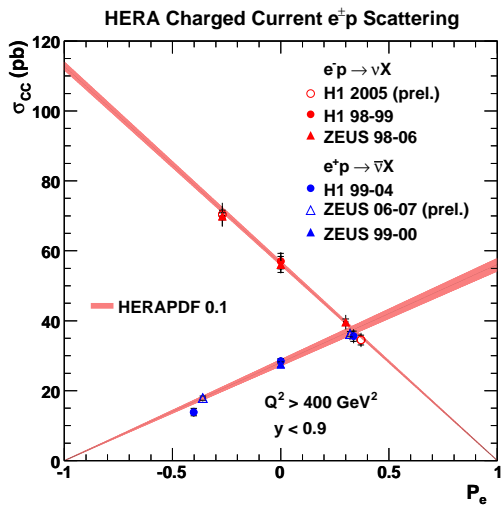,width=7.2cm}}
\vspace{-2.0mm}
\caption{\label{f:cc}  {\it Total cross sections for $e^-p$ and $e^+p$
               CC DIS as a function of longitudinal electron (positron) 
              beam polarization. \hspace{\fill}
 }}
\vspace{-2.0mm}
\end{figure}

The total CC cross sections have been measured by H1~\cite{H1_CC} and 
ZEUS~\cite{ZEUS_NCCC} as a function
of the lepton beam polarization $P_e$ in the
common phase space $Q^2 > 400$\,GeV$^2$ and $y<0.9$. 
In Fig.~\ref{f:cc} the results are compared with SM predictions 
based on CTEQ6D, MRST 2004 and HERAPDF0.1 fits. 
The linear dependence of the CC cross sections on $P_e$ 
is expected as the $W$ boson interacts only with $e_L^-$ and $e_R^+$.
A straight line fit to these cross sections is sensitive to exotic 
right-handed and left-handed current additions to the SM Lagrangian. 
Assuming SM couplings and a massless right-handed neutrino,
the existence of charged currents involving right-handed fermions
mediated by a boson of mass below 208\,GeV is excluded
at 95\% C.L.~\cite{H1WR}.
%
%

\section{Combined Electroweak-QCD Fits}

\begin{center}
\begin{figure}[h]
  \unitlength 1cm
  \begin{minipage}{12cm}
    \begin{minipage}{5.9cm}
      \begin{picture}(5.9,5.9)
        \leavevmode
        \epsfig{file=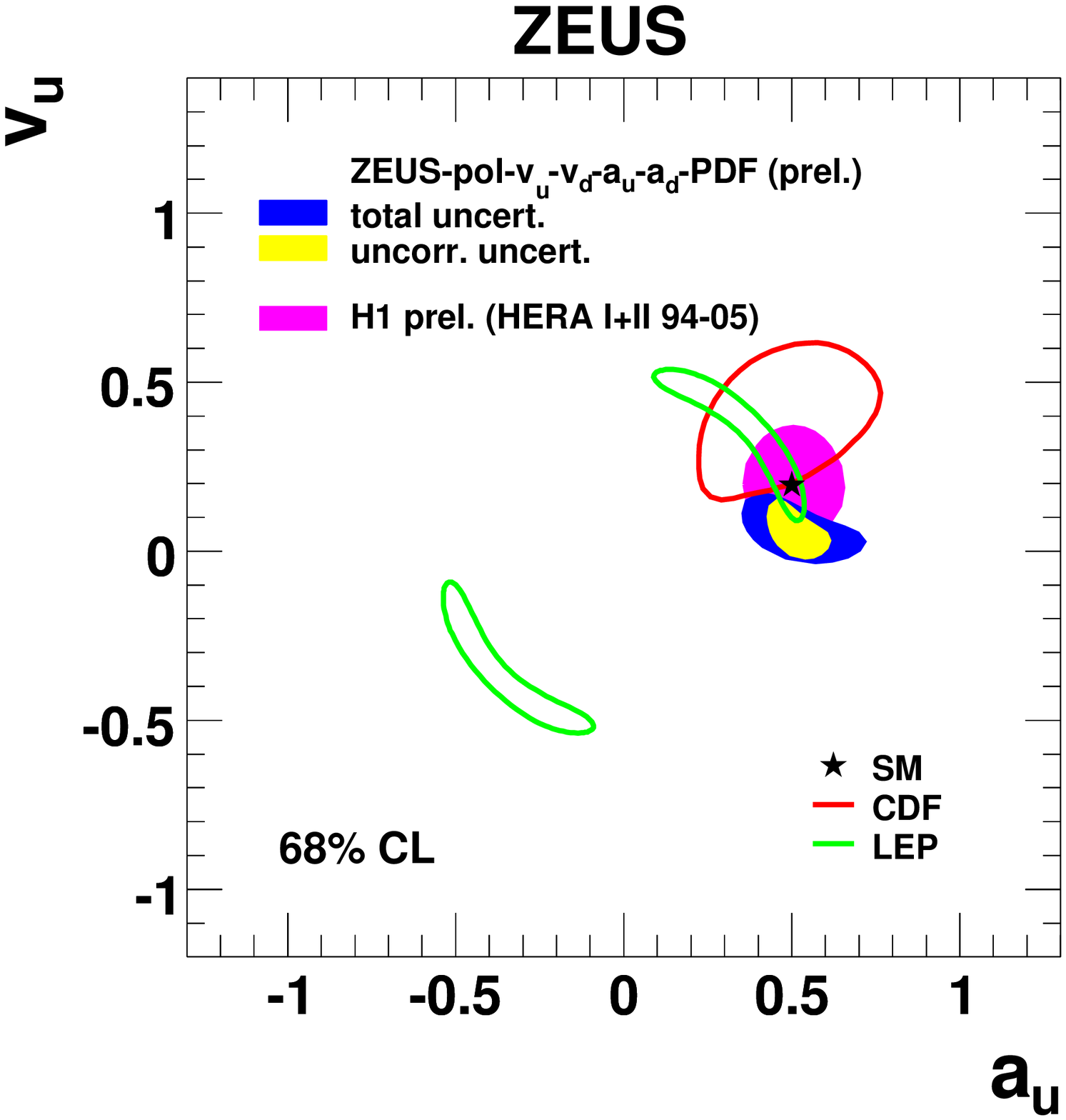,width=0.95\textwidth}
      \end{picture}
    \end{minipage}
    \hfill
    \begin{minipage}{5.9cm}
    \vspace{5.0mm}
      \begin{picture}(5.9,5.9)
          \leavevmode
        \epsfig{file=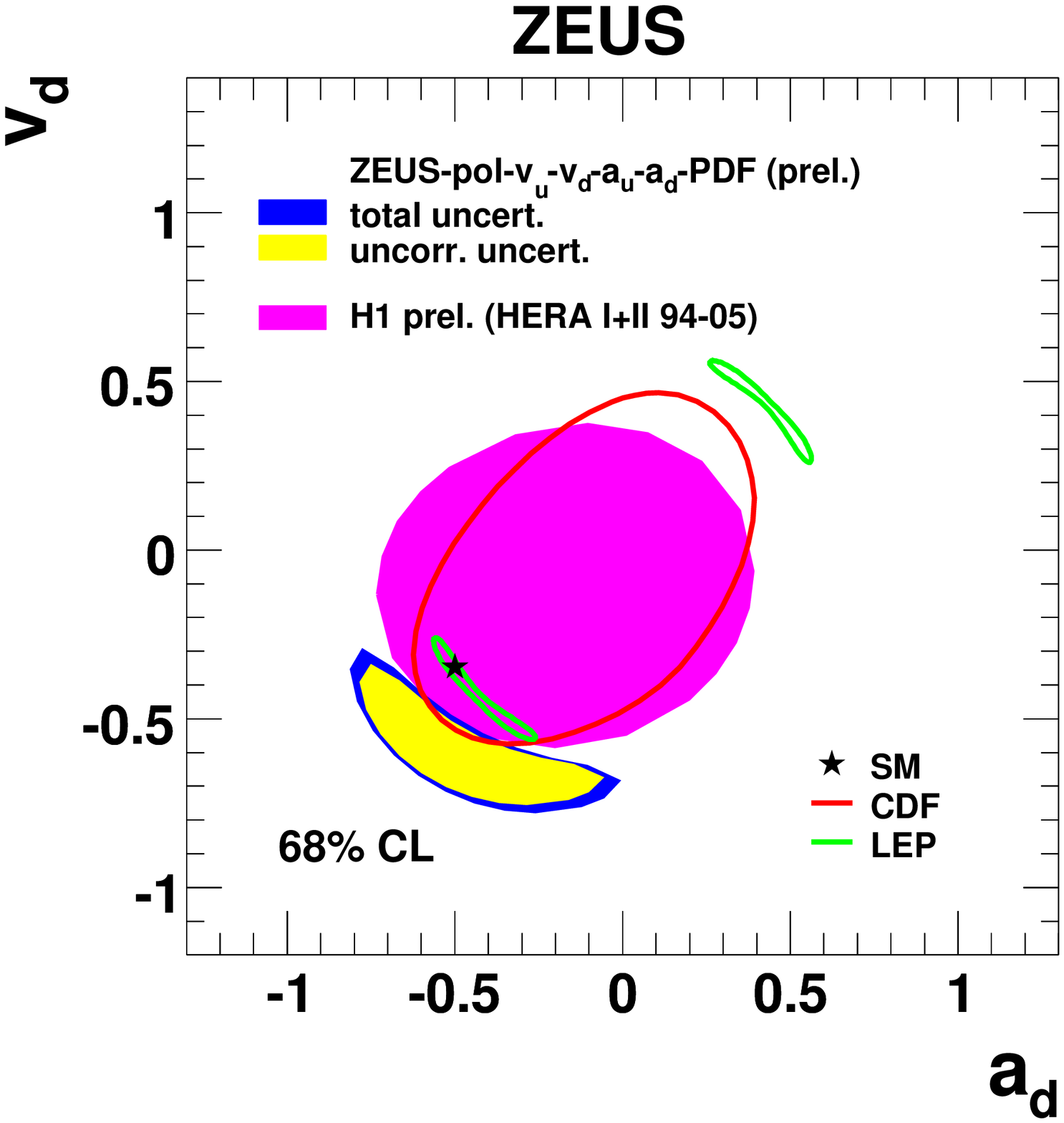,width=0.95\textwidth}
      \end{picture}
      \begin{center}
      \end{center}
    \end{minipage}
  \end{minipage}
\vspace{-10.0mm}
 \caption{ {\it Contour plots of the 68\%\,C.L. limits on the electroweak
                couplings of the quarks to the $Z^0$ boson. 
                Left side $v_u$ vs.\,$a_u$ (u quarks). 
                Right side $v_d$ vs.\,$a_d$ (d quarks).  \hspace{\fill}
	 }} 
 \label{f:couplings}
\end{figure}
\end{center}

The NC cross sections provide information on the quark couplings to the 
$Z^0$ boson. For the HERA kinematic regime, the axial $(a_q)$ and 
vector $(v_q)$ coupling constants are dominant in the unpolarized
$xF_3^0$ and polarized $F_2^P$ structure functions, respectively.
These electroweak parameters can be fitted simultaneously with the 
PDF parameters to perform a model independent extraction. 
The HERA~\cite{H1_EWQCD,ZEUS_EWQCD}
 results are shown in Fig.~\ref{f:couplings} and compared
to LEP and CDF results~\cite{LEP,CDF}. 

\section{Isolated Lepton Events with Missing {\boldmath $P_T$}}

A search for events with high transverse  energy isolated leptons 
(electrons or muons) 
and missing transverse momentum has been performed by the H1
 and ZEUS collaborations~\cite{H1ZEUS_iso}
using the full data sets. 
In general, the observed events yields are in good agreement with 
the SM model predictions, which is dominated by $W$ production. 
An excess at high $P_T^X$, $P_T^X >25$\,GeV, is observed by H1 in the 
$e^+ p$ data sample, which is not observed by ZEUS. 
A small excess remains in the common analysis~\cite{H1ZEUS_iso}: the 
number of observed events with $P_T^X >$ 25\,GeV is 23, compared to
$14.0\pm1.9$ expected.

\begin{figure} [h]
\centerline{\epsfig{figure=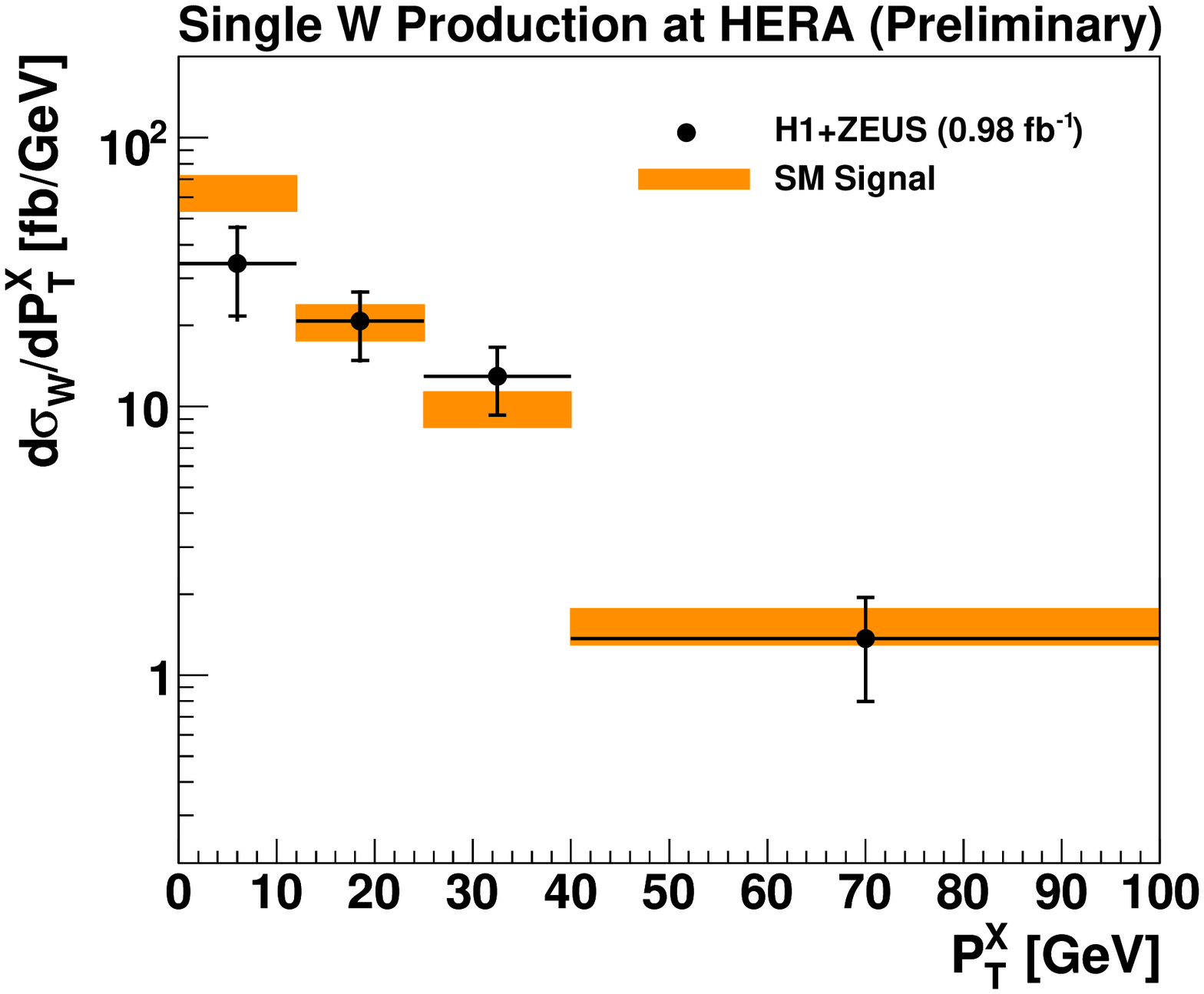,width=7.5cm}}
\vspace{-5.0mm}
\caption{\label{f:singleW}  {\it  The differential single $W$ cross section
 plotted as a function $P_T^X$.  \hspace{\fill}
 }}
\vspace{-2.0mm}
\end{figure}

The measured single $W$ production cross section is shown in
Fig.~\ref{f:singleW}. 
The total cross section of this process is measured as 
$1.07\pm0.16$(stat.)$\pm0.08$(sys.)\,pb, in agreement with the SM 
prediction of $1.26\pm0.19$\,pb. 

%

\section{Multi-Lepton Production}

The production of multileptons (electron or muon)  at high
transverse momenta has been studied by the  H1 and 
and ZEUS collaborations~\cite{H1ZEUS_multi} 
using the full $e^\pm p$ 
data sample. 
The yields of di- and tri-lepton events are in good agreement with SM 
predictions.
Distributions of the invariant mass $M_{12}$ of the two highest $P_T$ leptons
and of the scalar sum of the lepton transverse momenta $\sum P_T$
are in good agreement with the SM expectations.
Events are observed in $ee, e\mu, eee$ and $e\mu\mu$ topologies
with invariant masses $M_{12}$ above 100 GeV, where the SM prediction is low.
Both experiments observe high mass and high $\sum P_T$ events in 
$e^+ p$ collisions only, while, for comparable SM expectations, 
none are observed in $e^-p$ collisions. 
In the combined analysis 
seven events have a $\sum P_T >$ 100\,GeV, whereas the corresponding SM
expectation for $e^+ p$ collisions is $1.94 \pm 0.17$ 
events~\cite{H1ZEUS_multi} .

The total and differential cross sections for electron and muon pair
photo-production are measured in a restricted phase space dominated by 
photon-photon interactions 
and are found in good agreement 
with the SM expectations.

\section{Single-top Production}

Observables sensitive to flavor-changing neutral current (FCNC) interactions 
are particularly useful as probes for physics beyond the SM, since 
SM rates are very small due to the GIM mechanism. 
At the HERA collider, single-top production is a prime reaction to search
for FCNC, where the incoming lepton exchanges a $\gamma$ or $Z$ with an up-type
quark in the proton, producing a top quark in the final state. 
Deviations from the SM can be parameterized in terms of the coupling constants
$\kappa_{tu\gamma}$, $v_{tuZ}$~\cite{Han}. 

The studies performed by H1 and ZEUS considered top quark decays into
a $b$ quark and a $W$ boson with subsequent leptonic or 
hadronic decay of the $W$. The search is therefore based on 
a sample of events with isolated leptons and missing transverse 
momentum and a sample of multi-jet events. 
No evidence for single top quark production is observed. 
A 95\% C.L. limit on the anomalous coupling $\kappa_{tu\gamma}$, 
$\kappa_{tu\gamma} <0.18$ for H1~\cite{H1_stop} and 
$\kappa_{tu\gamma} <0.13$ for ZEUS~\cite{ZEUS_stop} 
is set for the scale of new physics of
$\Lambda \equiv m_{top} \equiv 175$\,GeV.

\section{Are Quarks elementary?}

A possible quark sub-structure can be detected by measuring the spatial
distribution of the quark charge. If the quark has a finite radius, the
cross section will decrease as the probes penetrates into it.
Deviations from the SM cross section are described by:
\begin{displaymath}
 \frac{d\sigma}{dQ^2} = \frac{d\sigma^{\rm SM}}{dQ^2}
        \left( 1 - \frac{R^2_q}{6}Q^2\right)^2
        \left( 1 - \frac{R^2_e}{6}Q^2\right)^2 ,
\end{displaymath}
where $R_e$ and $R_q$ are the root-mean-square radii of the electroweak
charge of the electron and the quark, respectively.

The high $Q^2$ ($Q^2 > 1\,000$\,GeV$^2$) neutral  current 
data sample has been used for this analysis. 
Assuming the electron to be point-like, the 95\% C.L.\,limit on the quark
radius is obtained as: 
$R_q < 0.74  \cdot 10^{-18}$\,m (H1~\cite{H1_NC}) and
$R_q < 0.63  \cdot 10^{-18}$\,m (ZEUS~\cite{ZEUS_rq}).

The ZEUS data set has also been used to derive limits on the mass scale 
parameter
in models with large extra dimensions and on the effective mass scale limits 
for contact-interaction model ranging from 3.8 to 8.9\,TeV~\cite{ZEUS_rq}.

\section{Search for Excited Fermions}

Excited fermions $(e^\star, \nu^\star $and $q^\star)$
would be a signature of  compositeness at the compositeness scale $\Lambda$. 
The cross section is proportional to the coupling 
constants $f$ and $f^{'}$~\cite{H1_ex_e}. 
All electroweak decays of excited fermions 
have been considered, including  final states from $Z$ or $W$ hadronic decays.
No evidence for excited  fermion production is found. 
Exclusion limits on $f/\Lambda$ at 95\% C.L. are determined by H1 as a
function of the mass of the excited fermions.
Assuming $f/\Lambda = 1/M_{f^\star}$, the following mass limits are
derived at 95\% C.L.: 
$M_{e^\star} > 272$\,GeV, 
$M_{\nu^\star} > 213$\,GeV and
$M_{q^\star} > 252$\,GeV (for $f_s = 0$)~\cite{H1_ex_e,H1_ex_q}.


\section{Leptoquarks}

A search for scalar and vector leptoquarks (LQ) coupling to first generation
fermions has been performed by the H1 collaboration using the full
HERA data set~\cite{H1_LQ}. 
Leptoquark decays into $eq$ and $\nu q$ were considered, where $q$
represents both quarks and anti-quarks. 
Such LQ decays lead to final states similar to those of DIS NC and CC
interactions at very high $Q^2$. 
No evidence for direct or indirect production of leptoquarks is found in data
samples with a large transverse momentum final state electron or with
large missing transverse momentum.
For each of the LQ species defined in the 
Buchm\"uller-R\"uckl-Wyler (BRW) model~\cite{BRW}, the
present analysis excludes a previously unexplored domain in the 
plane spanned by the mass of the leptoquark and its coupling to fermions.

As an example limits on the coupling $\lambda$ for 
$\tilde S_{1/2,L}$ and $S_{0,L}$ leptoquarks
are  shown in Fig.~\ref{f:LQ} 
as function  of the LQ  mass.

\begin{center}
\begin{figure}[h]
  \unitlength 1cm
  \begin{minipage}{12cm}
    \begin{minipage}{5.9cm}
      \hspace{-4mm}
      \begin{picture}(5.9,4)
        \leavevmode
        \epsfig{file=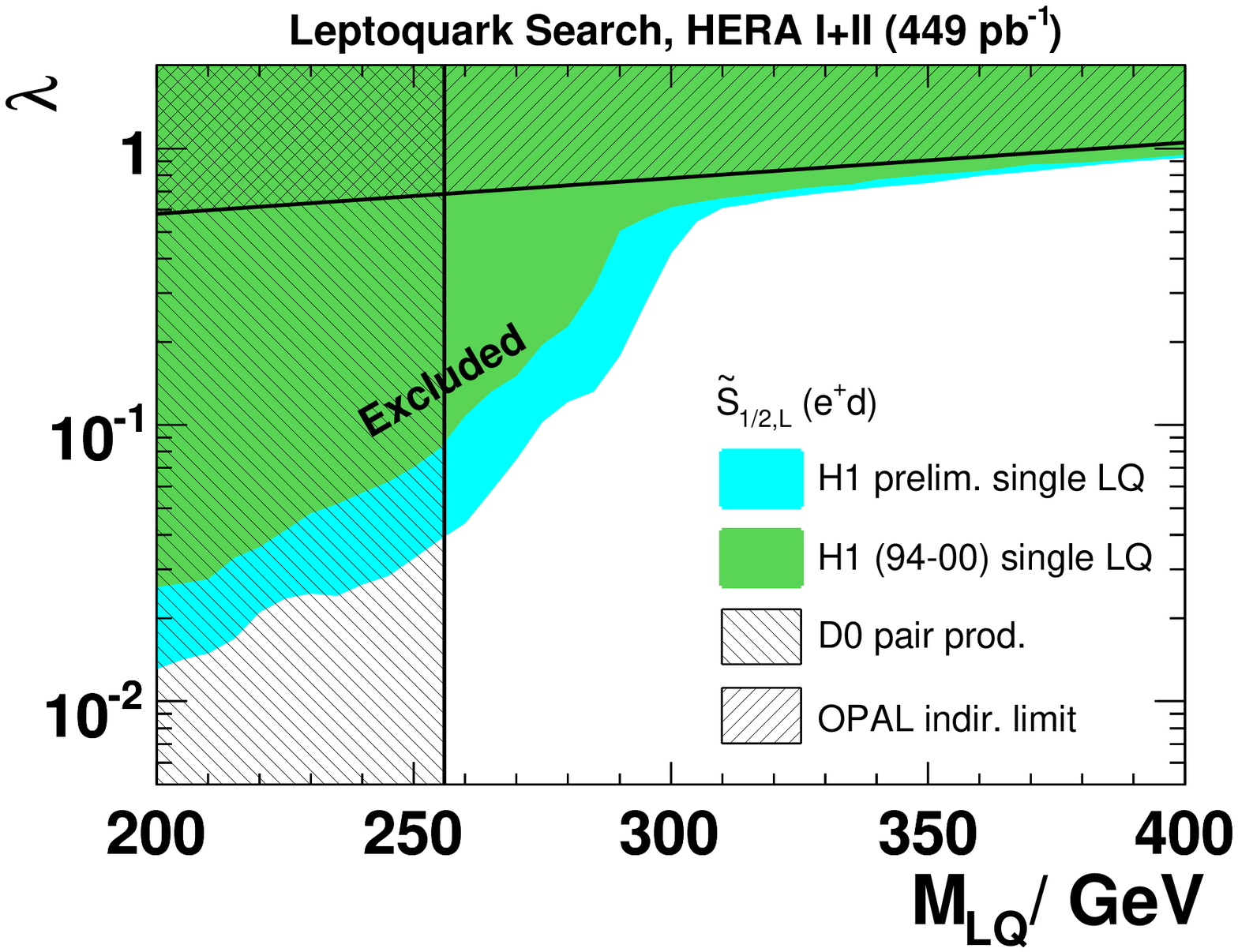,width=1.15\textwidth}
      \end{picture}
    \end{minipage}
    \hfill
    \begin{minipage}{5.9cm}
     \vspace{+5.0mm}
      \hspace{-1.5mm}
      \begin{picture}(5.9,4)
          \leavevmode
        \epsfig{file=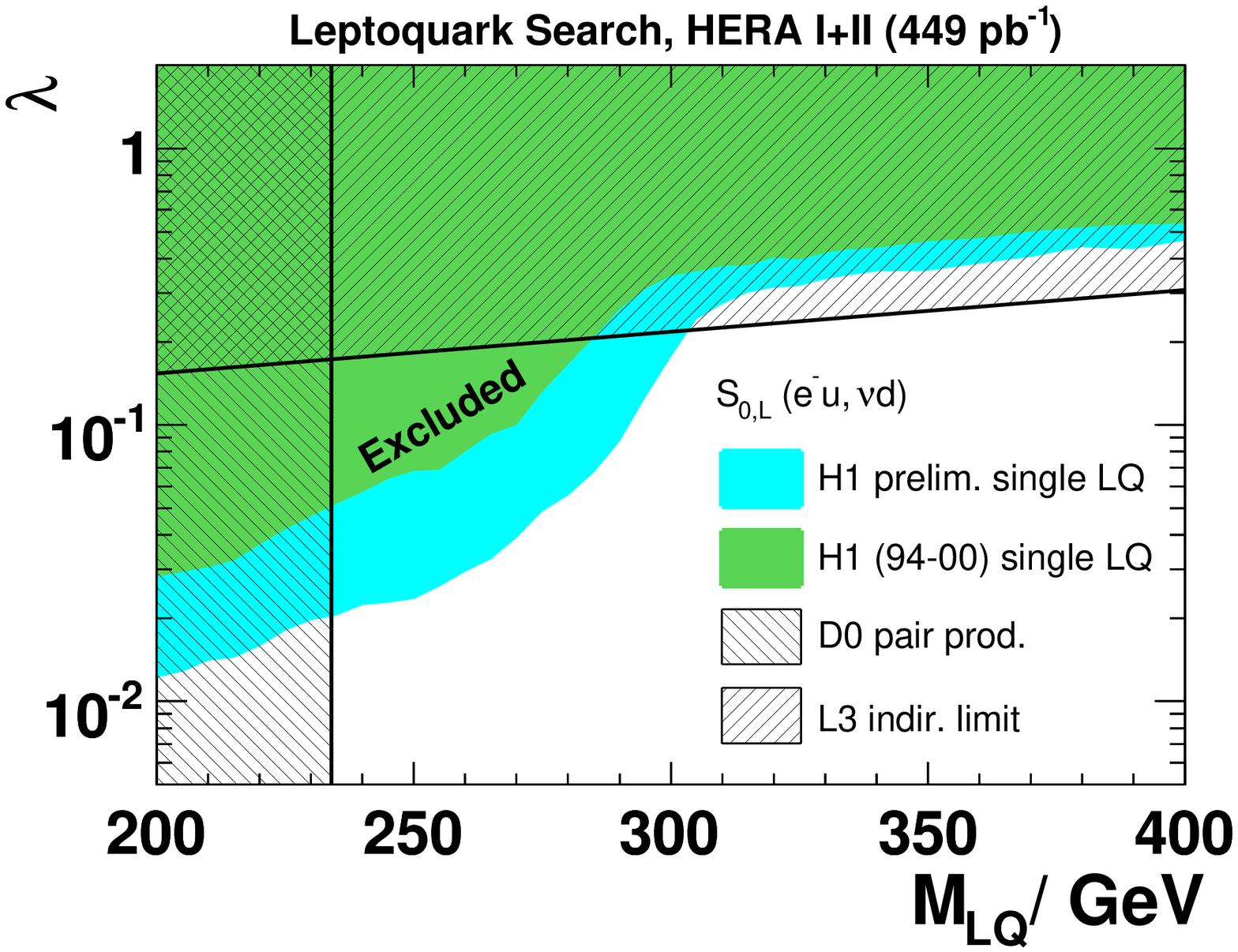,width=1.15\textwidth}
      \end{picture}
      \begin{center}
      \end{center}
    \end{minipage}
  \end{minipage}
\vspace{-8.0mm}

\caption{\label{f:LQ}  {\it Limits of the coupling constant $\lambda$
                for $\tilde S_{1/2,L}$ and $S_{0,L}$ leptoquarks as a
             function of their mass.   \hspace{\fill}
 }}
\end{figure}
\vspace{-5.0mm}
\end{center}

\section{General Searches for High - {\boldmath $P_T$} Phenomena}

H1 performed a model independent, generic search in final states with
at least two high-$P_T$ objects: electrons, muons, jets, photons or 
neutrinos ~\cite{H1_PT_searches}. 
The transverse momentum of these objects is required to be larger than 20\,GeV.
The events were classified according to their final states. Forty different
final states  were considered.  
In general, the events yields are in  good agreement with 
Standard Model expectations. No statistically significant 
deviation is observed.

\section*{Acknowledgments}

I would like to thank the organizers, in particular A.Studenikin, 
for the invitation to the conference. I am grateful to  M.Turcato 
and S.Schmitt for a critical reading of the manuscript.


\section*{References}
\begin{thebibliography}{99}


\bibitem{Cooper} A.M.~Cooper-Sarkar, R.C.E.~Devenish and  A.~De~Roeck,
                {\it Int.\ J.\ Mod.\ Phys.{}} {\bf A~13},~3385~(1998). 

\bibitem{H1_NC} H1 Collaboration, prelim-07-141.

\bibitem{H1_CC} H1 Collaboration, prelim-06-041.

\bibitem{ZEUS_NCCC} ZEUS Collaboration, 
                 ZEUS-prel-09-001.

\bibitem{H1WR} A.~Aktas et al., H1 Collaboration,
                {\it Phys.\ Lett.} {\bf B634}, 173  ~(2006). 


\bibitem{ZEUS_xF3} S.Chekanov et al., ZEUS Collaboration, 
                {\it Eur.\ Phys.\ J.} {\bf C 62}, 625 (2009).

\bibitem{H1_ZEUS_PolA} H1 and ZEUS Collaborations, H1 prelim-06-142 and 
                     ZEUS prel-06-022. 

\bibitem{H1_EWQCD} A.~Aktas et al., H1 Collaboration,
                {\it Phys.\ Lett.} {\bf B632}, 35  ~(2006). 

\bibitem{ZEUS_EWQCD} S.Chekanov et al., ZEUS Collaboration, 
                 ZEUS-prel-07-027.

\bibitem{CDF} D.~Acosta et al., CDF Collaboration,
                {\it Phys.\ Lett.} {\bf B634}, 173  ~(2006). 

\bibitem{LEP} LEP Electroweak Working Group, 
                {\it Phys.\ Rep.} {\bf 427}, 257  ~(2006). 


\bibitem{H1ZEUS_iso} H1 and ZEUS Collaborations,
                   H1 prelim-09-161, ZEUS-prel-09-044.
 
\bibitem{H1ZEUS_multi}  H1 and ZEUS Collaborations,  
                {\it JHEP10} 013 (2009).

\bibitem{Han} T.~Han, J.L.~Hewett,
                {\it Phys.\ Rev.} {\bf D60},{074015} (1999). 

\bibitem{H1_stop} F.D.Aaron et al., H1 Collaboration, DESY 09-050 (2009).

\bibitem{ZEUS_stop} ZEUS Collaboration, ZEUS-prel-09-009.

\bibitem{ZEUS_rq} ZEUS Collaboration, ZEUS-prel-09-013.

\bibitem{H1_ex_e} F.D.Aaron et al., H1 Collaboration,  
                {\it Phys.\ Lett.} {\bf B678}, 335 (2009).

\bibitem{H1_ex_q} F.D.Aaron et al., H1 Collaboration,  
                DESY 09-040, arXiv:0904.3392v2 (2009).

\bibitem{H1_LQ} H1 Collaboration,
                   H1 prelim-07-164, (2007)

\bibitem{BRW} W.Buchm\"uller, R.R\"uckl and D.Wyler, Phys.Lett. {\bf B191}, 
        442 (1987).   Erratum in  Phys.Lett. {\bf B448}, 320 (1999).

\bibitem{H1_PT_searches} F.D.Aaron et al., H1 Collaboration,  
                {\it Phys.\ Lett.} {\bf B 674}, 257 (2009).

\end{thebibliography}
\end{document}